\documentclass[letterpaper, 10pt, conference, twocolumn]{IEEEtran}
\usepackage{fancyhdr}
\usepackage{amsmath,epsfig}
\usepackage{threeparttable}
\usepackage{epsf,epsfig}
\usepackage{amsmath}
\usepackage{amssymb}
\usepackage{amsfonts}
\usepackage[noadjust]{cite}
\usepackage{dsfont}
\usepackage{subfigure}

\begin{document}
\newtheorem{theorem}{\it Theorem}
\newtheorem{acknowledgement}[theorem]{Acknowledgement}
\newtheorem{axiom}[theorem]{Axiom}
\newtheorem{case}[theorem]{Case}
\newtheorem{claim}[theorem]{Claim}
\newtheorem{conclusion}[theorem]{Conclusion}
\newtheorem{condition}[theorem]{Condition}
\newtheorem{conjecture}[theorem]{Conjecture}
\newtheorem{criterion}[theorem]{Criterion}
\newtheorem{definition}[theorem]{Definition}
\newtheorem{example}[theorem]{Example}
\newtheorem{exercise}[theorem]{Exercise}
\newtheorem{lemma}{Lemma}
\newtheorem{corollary}{Corollary}
\newtheorem{notation}[theorem]{Notation}
\newtheorem{problem}[theorem]{Problem}
\newtheorem{proposition}{Proposition}
\newtheorem{solution}[theorem]{Solution}
\newtheorem{summary}[theorem]{Summary}
\newtheorem{assumption}{Assumption}
\newtheorem{examp}{\bf Example}
\newtheorem{probform}{\bf Problem}
\def\remark{{\noindent \bf Remark:\hspace{0.5em}}}

\def\qed{$\Box$}
\def\QED{\mbox{\phantom{m}}\nolinebreak\hfill$\,\Box$}
\def\proof{\noindent{\emph{Proof:} }}
\def\poof{\noindent{\emph{Sketch of Proof:} }}
\def
\endproof{\hspace*{\fill}~\qed
\par
\endtrivlist\unskip}
\def\endproof{\hspace*{\fill}~\qed\par\endtrivlist\vskip3pt}

\def\E{\mathbf{E}}
\def\eps{\varepsilon}
\def\phi{\varphi}
\def\Lsp{{\boldsymbol L}}
\def\Bsp{{\boldsymbol B}}
\def\lsp{{\boldsymbol\ell}}
\def\Ltsp{{\Lsp^2}}
\def\Lpsp{{\Lsp^p}}
\def\Linsp{{\Lsp^{\infty}}}
\def\LtR{{\Lsp^2(\Rst)}}
\def\ltZ{{\lsp^2(\Zst)}}
\def\ltsp{{\lsp^2}}
\def\ltZt{{\lsp^2(\Zst^{2})}}
\def\ninN{{n{\in}\Nst}}
\def\oh{{\frac{1}{2}}}
\def\grass{{\cal G}}
\def\ord{{\cal O}}
\def\dist{{d_G}}
\def\conj#1{{\overline#1}}
\def\ntoinf{{n \rightarrow \infty }}
\def\toinf{{\rightarrow \infty }}
\def\tozero{{\rightarrow 0 }}
\def\trace{{\operatorname{trace}}}
\def\ord{{\cal O}}
\def\UU{{\cal U}}
\def\rank{{\operatorname{rank}}}
\def\acos{{\operatorname{acos}}}

\def\SINR{\mathrm{SINR}}
\def\SNR{\mathrm{SNR}}
\def\SIR{\mathrm{SIR}}

\setcounter{page}{1}

% Definitions
\newcommand{\eref}[1]{(\ref{#1})}
\newcommand{\fig}[1]{Fig.\ \ref{#1}}

% Bold lowercase
\def\bydef{:=}
\def\ba{{\mathbf{a}}}
\def\bb{{\mathbf{b}}}
\def\bc{{\mathbf{c}}}
\def\bd{{\mathbf{d}}}
\def\bee{{\mathbf{e}}}
\def\bff{{\mathbf{f}}}
\def\bg{{\mathbf{g}}}
\def\bh{{\mathbf{h}}}
\def\bi{{\mathbf{i}}}
\def\bj{{\mathbf{j}}}
\def\bk{{\mathbf{k}}}
\def\bl{{\mathbf{l}}}
\def\bm{{\mathbf{m}}}
\def\bn{{\mathbf{n}}}
\def\bo{{\mathbf{o}}}
\def\bp{{\mathbf{p}}}
\def\bq{{\mathbf{q}}}
\def\br{{\mathbf{r}}}
\def\bs{{\mathbf{s}}}
\def\bt{{\mathbf{t}}}
\def\bu{{\mathbf{u}}}
\def\bv{{\mathbf{v}}}
\def\bw{{\mathbf{w}}}
\def\bx{{\mathbf{x}}}
\def\by{{\mathbf{y}}}
\def\bz{{\mathbf{z}}}
\def\b0{{\mathbf{0}}}

% Bold capital letters
\def\bA{{\mathbf{A}}}
\def\bB{{\mathbf{B}}}
\def\bC{{\mathbf{C}}}
\def\bD{{\mathbf{D}}}
\def\bE{{\mathbf{E}}}
\def\bF{{\mathbf{F}}}
\def\bG{{\mathbf{G}}}
\def\bH{{\mathbf{H}}}
\def\bI{{\mathbf{I}}}
\def\bJ{{\mathbf{J}}}
\def\bK{{\mathbf{K}}}
\def\bL{{\mathbf{L}}}
\def\bM{{\mathbf{M}}}
\def\bN{{\mathbf{N}}}
\def\bO{{\mathbf{O}}}
\def\bP{{\mathbf{P}}}
\def\bQ{{\mathbf{Q}}}
\def\bR{{\mathbf{R}}}
\def\bS{{\mathbf{S}}}
\def\bT{{\mathbf{T}}}
\def\bU{{\mathbf{U}}}
\def\bV{{\mathbf{V}}}
\def\bW{{\mathbf{W}}}
\def\bX{{\mathbf{X}}}
\def\bY{{\mathbf{Y}}}
\def\bZ{{\mathbf{Z}}}

% mathbb Bold capital letters
\def\mA{{\mathbb{A}}}
\def\mB{{\mathbb{B}}}
\def\mC{{\mathbb{C}}}
\def\mD{{\mathbb{D}}}
\def\mE{{\mathbb{E}}}
\def\mF{{\mathbb{F}}}
\def\mG{{\mathbb{G}}}
\def\mH{{\mathbb{H}}}
\def\mI{{\mathbb{I}}}
\def\mJ{{\mathbb{J}}}
\def\mK{{\mathbb{K}}}
\def\mL{{\mathbb{L}}}
\def\mM{{\mathbb{M}}}
\def\mN{{\mathbb{N}}}
\def\mO{{\mathbb{O}}}
\def\mP{{\mathbb{P}}}
\def\mQ{{\mathbb{Q}}}
\def\mR{{\mathbb{R}}}
\def\mS{{\mathbb{S}}}
\def\mT{{\mathbb{T}}}
\def\mU{{\mathbb{U}}}
\def\mV{{\mathbb{V}}}
\def\mW{{\mathbb{W}}}
\def\mX{{\mathbb{X}}}
\def\mY{{\mathbb{Y}}}
\def\mZ{{\mathbb{Z}}}

% Caligraphic capital letters
\def\cA{\mathcal{A}}
\def\cB{\mathcal{B}}
\def\cC{\mathcal{C}}
\def\cD{\mathcal{D}}
\def\cE{\mathcal{E}}
\def\cF{\mathcal{F}}
\def\cG{\mathcal{G}}
\def\cH{\mathcal{H}}
\def\cI{\mathcal{I}}
\def\cJ{\mathcal{J}}
\def\cK{\mathcal{K}}
\def\cL{\mathcal{L}}
\def\cM{\mathcal{M}}
\def\cN{\mathcal{N}}
\def\cO{\mathcal{O}}
\def\cP{\mathcal{P}}
\def\cQ{\mathcal{Q}}
\def\cR{\mathcal{R}}
\def\cS{\mathcal{S}}
\def\cT{\mathcal{T}}
\def\cU{\mathcal{U}}
\def\cV{\mathcal{V}}
\def\cW{\mathcal{W}}
\def\cX{\mathcal{X}}
\def\cY{\mathcal{Y}}
\def\cZ{\mathcal{Z}}
\def\cd{\mathcal{d}}
\def\Mt{M_{t}}
\def\Mr{M_{r}}
%% my defs
\def\O{\Omega_{M_{t}}}
\newcommand{\figref}[1]{{Fig.}~\ref{#1}}
\newcommand{\tabref}[1]{{Table}~\ref{#1}}

%% From Kaibin
\newcommand{\var}{\mathrm{Var}}
\newcommand{\fb}{\tx{fb}}
\newcommand{\nf}{\tx{nf}}
\newcommand{\BC}{\tx{(bc)}}
\newcommand{\MAC}{\tx{(mac)}}
\newcommand{\Pout}{P_{\tx{out}}}
\newcommand{\nnn}{\nn\\}
\newcommand{\FB}{\tx{FB}}
\newcommand{\TX}{\tx{TX}}
\newcommand{\RX}{\tx{RX}}
\renewcommand{\mod}{\tx{mod}}
\newcommand{\m}[1]{\mathbf{#1}}
\newcommand{\td}[1]{\tilde{#1}}
\newcommand{\sbf}[1]{\scriptsize{\textbf{#1}}}
\newcommand{\stxt}[1]{\scriptsize{\textrm{#1}}}
\newcommand{\suml}[2]{\sum\limits_{#1}^{#2}}
\newcommand{\sumlk}{\sum\limits_{k=0}^{K-1}}
\newcommand{\eqhsp}{\hspace{10 pt}}
\newcommand{\tx}[1]{\texttt{#1}}
\newcommand{\Hz}{\ \tx{Hz}}
\newcommand{\sinc}{\tx{sinc}}
\newcommand{\tr}{\mathrm{tr}}
\newcommand{\diag}{\mathrm{diag}}
\newcommand{\MAI}{\tx{MAI}}
\newcommand{\ISI}{\tx{ISI}}
\newcommand{\IBI}{\tx{IBI}}
\newcommand{\CN}{\tx{CN}}
\newcommand{\CP}{\tx{CP}}
\newcommand{\ZP}{\tx{ZP}}
\newcommand{\ZF}{\tx{ZF}}
\newcommand{\SP}{\tx{SP}}
\newcommand{\MMSE}{\tx{MMSE}}
\newcommand{\MINF}{\tx{MINF}}
\newcommand{\RC}{\tx{MP}}
\newcommand{\MBER}{\tx{MBER}}
\newcommand{\MSNR}{\tx{MSNR}}
\newcommand{\MCAP}{\tx{MCAP}}
\newcommand{\vol}{\tx{vol}}
\newcommand{\ah}{\hat{g}}
\newcommand{\tg}{\tilde{g}}
\newcommand{\teta}{\tilde{\eta}}
\newcommand{\heta}{\hat{\eta}}
\newcommand{\uh}{\m{\hat{s}}}
\newcommand{\eh}{\m{\hat{\eta}}}
\newcommand{\hv}{\m{h}}
\newcommand{\hh}{\m{\hat{h}}}
\newcommand{\Po}{P_{\mathrm{out}}}
\newcommand{\Poh}{\hat{P}_{\mathrm{out}}}
\newcommand{\Ph}{\hat{\gamma}}
\newcommand{\mat}[1]{\begin{matrix}#1\end{matrix}}
\newcommand{\ud}{^{\dagger}}
\newcommand{\C}{\mathcal{C}}
\newcommand{\nn}{\nonumber}
\newcommand{\nInf}{U\rightarrow \infty}

\title{Spatial Interference Cancelation for Mobile Ad Hoc Networks: Perfect CSI}

\author{\authorblockN{Kaibin Huang, Jeffrey G. Andrews,  Robert W. Heath, Jr.}
\authorblockA{
Wireless Networking \& Communications Group\\
Department of Electrical \& Computer Engineering\\
The University of Texas at Austin, Austin, TX 78712-0240 \\
Email: khuang@mail.utexas.edu, \{jandrews, rheath\}@ece.utexas.edu\vspace{-100pt}}
\and
\authorblockN{Dongning Guo, Randall A. Berry }
\authorblockA{
Department of Electrical \& Computer Science\\
Northwestern University, Evanston, IL 60208\\
Email: dguo@northwestern.edu\\
\hspace{52pt}rberry@ece.northwestern.edu}}

\maketitle

\begin{abstract}
Interference between nodes directly limits the capacity of mobile ad hoc networks. This paper focuses on spatial interference cancelation with perfect channel state information (CSI), and analyzes the corresponding network capacity.  Specifically,  by using multiple antennas, zero-forcing beamforming is applied at each receiver for canceling the strongest interferers. Given spatial interference cancelation, the network transmission capacity is analyzed in this paper, which is defined as the maximum transmitting node density under constraints on outage and the signal-to-interference-noise ratio. Assuming the Poisson distribution for the locations of network nodes and spatially i.i.d. Rayleigh fading channels, mathematical tools from stochastic geometry are applied for deriving scaling laws for transmission capacity. Specifically, for small target outage probability, transmission capacity is proved to increase following a power law, where the exponent is the inverse of the size of antenna array or larger depending on the pass loss exponent. As shown by simulations, spatial interference cancelation increases transmission capacity by an order of magnitude or more even if only one extra antenna is added to each node.

\end{abstract}
\section{Introduction}\label{Section:Intro}
In a mobile ad hoc network (MANET), the mutual interference between nodes limits  throughput for peer-to-peer communication over the network. In this paper, zero-forcing beamforming are applied at receivers for canceling  interference from the strongest interferers. Thereby the number of successful communication links per unit area, called network \emph{transmission capacity} \cite{WeberAndrews:TransCapWlssAdHocNetwkOutage:2005}, increases significantly.

For a multi-hop ad hoc network, the notion of \emph{transport capacity} was introduced and analyzed in  \cite{GuptaKumar:CapWlssNetwk:2000}, which started a series of related studies (see e.g. \cite{FrancTse:CloseGapWlssNetwkPercolatnTheo:2007, Jovicic:UppBndTranpCapWlssNetwk:2004, AyferTse:HierCoopOptimCapAdHocNetwk:2006}). Prior results on transport capacity typically focus on scaling laws of network throughput with an asymptotically large number of nodes. Such asymptotic results may differ significantly from the actual throughput of finite-size networks, and thus have limited practical applications.

For single-hop ad hoc networks, the \emph{transmission capacity} metric introduced in \cite{WeberAndrews:TransCapAdHocNetwkDistSch:2006} is defined as the maximum number of successful communication links  per unit area under signal-to-interference-noise ratio (SINR) and outage constraints. By modeling network nodes as a Poisson point process,  this capacity measure enjoys more accurate analysis and easier computability compared with transport capacity \cite{WeberAndrews:TransCapWlssAdHocNetwkOutage:2005, WeberAndrews:TransCapAdHocNetwkDistSch:2006}. Transmission capacity has been used to make tractable analysis of various issues related to MANETs such as opportunistic transmission \cite{WeberAndrews:TransCapAdHocNetwkDistSch:2006}, successive interference cancelation (SIC) \cite{WeberAndrews:TransCapWlssAdHocNetwkSIC:2005} and multi-antenna transmission \cite{AndrewJeff:CapacityScalingSpatialDiversity:2006}.

Interference directly limits the throughput of MANETs. The optimal approach for reducing interference in MANETs is called \emph{interference alignment}, which achieves the number of degrees of freedom equal to half of the number of interference links \cite{CadJafar:InterfAlignment:2007}. Nevertheless, this approach appears daunting because it requires nodes to employ jointly designed precoders and obtain perfect channel sate information (CSI) of interference channels. An alternative approach is to use physical layer techniques of the multiuser detection family for suppressing interference \cite{VerBook}. These algorithms, however, appear to be very sensitive to residual interference due to imperfect interference cancelation, and the near-far problem.

This paper considers zero-forcing beamforming for interference cancelation in multi-antenna MANETs with single-stream data links. Therefore, the spatial degrees of freedoms created by multi-antennas are dedicated for interference cancelation. Recently,
beamforming or directional antennas have been integrated with the medium access control (MAC) protocols for MANETs to improve network spatial reuse efficiency (see e.g. \cite{Mundarath:CrossLayerAdaptiveAntAdHocNetworks:2007, Park:SPACE_MAC_MIMO:2005, Ramanathan:AdHocNetworkDirectAntennas:2005, Singh:SpatialReuseInterfCancel:2005}). Most prior work focuses on designing MAC protocols and relies on simulations. In \cite{AndrewJeff:CapacityScalingSpatialDiversity:2006}, the transmission capacity for multi-antenna MANETs is analyzed, where interference is treated as noise and suppressed by averaging through beamforming. There still lacks analysis of the effects of spatial interference cancelation on the transmission capacity of MANETs, which is addressed in this paper. Another important issue, namely how CSI inaccuracy affects the transmission capacity of MANETs, is investigated in a sperate paper \cite{Huang:SpatialInterfCancel:ImpCSI:Globecom08}.

The contributions of this paper are summarized as follows. This paper targets a MANET with single-stream data links and perfect synchronization between nodes. First, assuming perfect CSI, zero-forcing beamforming is applied for canceling interference at receivers and thereby increasing network transmission capacity. Second, based on the Poisson assumption on transmitting-node locations and the spatially i.i.d. Rayleigh fading channel model, bounds on the signal-to-interference ratio (SIR) outage probability are derived. Third, the scaling laws for transmission capacity are derived for asymptotically small target outage probability. Specifically, the asymptotic transmission capacity grows following a power law, where the exponent is the inverse of the antenna-array size or larger depending on the pass-loss exponent.

\section{Network and Channel Models } \label{Section:Sys}

\subsection{Network Model}\label{Section:NetworkModel}
In this paper, the locations of potential transmitting nodes in the mobile ad hoc network, including both active and inactive transmitters,  are modeled as a Poisson point process following the common approach in the literature \cite{WeberAndrews:TransCapWlssAdHocNetwkOutage:2005, WeberAndrews:TransCapWlssAdHocNetwkSIC:2005, WeberAndrews:TransCapAdHocNetwkDistSch:2006, AndrewJeff:CapacityScalingSpatialDiversity:2006}. Specifically, the positions of the potential transmitters form a homogeneous Poisson point process with the density denoted by $\lambda_o$. Each transmitter transmits with fixed probability denoted by $P_t$. Let $T_n$ denote the coordinate of the $i$th transmitter on the 2-D plane. The set $\Phi = \{T_n\}$ is also a homogeneous Poisson point process but with a smaller density $\lambda = P_t\lambda_o$ \cite{Kingman93:PoissonProc}. Each transmitter is associated with a receiver located at a fixed distance denoted as $d$.

Consider a typical receiver located at the origin, denoted as $R_0$, and hence $|T_0|=d$. This location constraint of $T_0$ does not compromise the generality since the transmitting node process $\Phi$ is translation invariant. Furthermore, according to Slivnyak's theorem \cite{Kingman93:PoissonProc}, other transmitters, namely $\Phi/\{T_0\}$, remain as a homogeneous Poisson point process with the same node density $\lambda$ .

The ad hoc network is assumed to be interference limited and thus noise is neglected for simplicity. Consequently, the reliability of data packets received by the node $R_0$ is determined by the SIR. Moreover, we assume that each data link in the network has a single stream, and communications between nodes are perfectly synchronized. Let $S$ denote the random channel power for the link from $T_0$ to $R_0$, and the function $I(T_n)$ gives the power of interference from the $T_n$ to $R_0$. Thus, assuming uniform data transmission power for all transmitters, the SIR at $R_0$ is given as $\SIR_0 = \frac{S}{\sum_{T_n\in\Phi/\{T_0\}}I(T_n)}$.
To simplify notation, $I(T_n)$ is denoted as $I_n$ in the sequel. The correct decoding of received data packets requires the SIR to exceed a threshold $\theta$. In other words, the rate of information sent from a transmitter to a receiver is $\log_2(1+\theta)$ assuming Gaussian signaling. To support this information rate with high probability, the outage probability that $\SIR_0$ is below $\theta$ must be smaller than or equal to a given threshold $0<\epsilon\ll 1$, i.e.
\begin{equation}\label{Eq:Pout:Def}
\Pout(\lambda) = \Pr(\SIR_0\leq \theta) \leq \epsilon
\end{equation}
where $\Pout(\lambda)$ denotes the SIR outage probability.  Given $\epsilon$, $\Pout$ determines the transmission capacity defined as \cite{WeberAndrews:TransCapWlssAdHocNetwkOutage:2005}
\begin{equation}\label{Eq:TxCap}
    C(\epsilon) = (1-\epsilon)\lambda_{\epsilon}
\end{equation}
where $\Pout(\lambda_\epsilon) = \epsilon$.

\subsection{Channel Model}\label{Section:ChannelModel}
The channel model is characterized by narrow-band and flat fading. Each node in the network is equipped with $L$ antennas. Consequently, there exists a $L\times L$ multiple-input-multiple-output (MIMO) channel between every pair of nodes. Each MIMO channel consists of path-loss and spatially i.i.d. small fading components, corresponding to rich scattering. Specifically, the channel from a node $T_n$ to the typical receiving node $R_0$ is $\bH_n = d_n^{-\alpha/2}\bG_n$. The factor $d_n^{-\alpha/2}$ represents path-loss, where $d_n = |T_n|$ and  $\alpha > 2$  is the path-loss exponent. The other factor of $\bH_n$, $\bG_n$, models spatially i.i.d. Rayleigh fading, and hence $\bG$ is a $L\times L$ matrix of i.i.d. $\mathcal{CN}(0,1)$ components. The above channel model simplifies analysis in this paper. Finally, given single-stream data links, beamforming is applied at each transmitter and receiver.

\section{Spatial Interference Cancelation}\label{Section:InterfAvoid}
Assume  perfect CSI,  synchronization between nodes, and single-stream data links. Under these assumptions, spatial interference cancelation uses zero-forcing beamforming by following the procedure described in Section~\ref{Section:ZFBeam:Algo}. The effective network and channel models are presented in Section~\ref{Section:EffModel:PerfCSI}. Without loss of generality, the discussion focuses on the typical pair of nodes $T_0$ and $R_0$ (cf. Section~\ref{Section:NetworkModel}).

\subsection{Algorithms} \label{Section:ZFBeam:Algo}
The idea of spatial interference cancelation is to apply zero-forcing beamforming at $R_0$ for canceling interference from strong interferers. Let $\bff_n$ and $\bv_0$ denote the transmit beamformer at $T_n$ and the receive beamformer at $R_0$, respectively. From the perspective of $R_0$, the interference channel from $T_n$ ($n\neq 0$)  appears as an effective channel vector $\bh_n = \bH_n\bff_n$. To facilitate our discussion, the indices of the interferers of $R_0$ are sorted according to their effective interference channel norms, namely $\|\bh_1\|\geq \|\bh_2\| \geq \cdots\geq \|\bh_L\|\cdots$. By zero-forcing beamforming, the beamforming vector $\bv_0$ of $R_0$ is constrained to be in the null space of the matrix $[\bh_1,\bh_2,\cdots, \bh_{L-1}]$. Thereby, the interference from $L-1$ strongest interferers to $R_0$ is canceled. Note that perfect CSI estimation  of $\bh_1,\bh_2,\cdots, \bh_{L-1}$ by $R_0$ is required to completely cancel the interference from $L-1$ strongest interferers.  CSI estimation at each receiver uses  pilot symbols broadcast by transmitters. The issue of CSI inaccuracy is addressed in \cite{Huang:SpatialInterfCancel:ImpCSI:Globecom08}. Next, an arbitrary transmit beamformer is applied at $T_0$, represented by $\bff_0$. By such beamforming, multiple transmit antennas  contribute no diversity gain.

To avoid deep fading due to the lack of diversity gain, opportunistic transmission is applied as in \cite{WeberAndrews:TransCapAdHocNetwkDistSch:2006}. Consequently, transmission at each transmitter is turned on only if the channel gain $S$ is above a threshold denoted by $\beta$, where $S = |\bv_0^\dagger\bH_0\bff_0|^2$. It follows that the transmission probability for each potential transmitter is $P_t = \Pr(S\geq \beta)$ (cf. Section~\ref{Section:NetworkModel}).

\subsection{Effective Channel and Network Models}\label{Section:EffModel:PerfCSI}
With perfect interference cancelation, $R_0$ receives interference only from the nodes $\{T_n\mid n\geq L\}$. Let $r_n$ and $I_n$ denote respectively the distance between $T_n$ and the origin, and the interference power from $T_n$ to $R_0$. Using this notation and based on the channel model in Section~\ref{Section:ChannelModel}, for $n\geq L$, $I_n = P_D|\bv_0^\dagger\bH_n\bff_n|^2= r_n^{-\alpha}|\bv_0^\dagger\bG_n\bff_n|^2$. Because both $\bff_n$ and $\bv_0$ are independent of $\bG_n$ and $\bG_n$ is an i.i.d. complex Gaussian matrix, the random variable $\rho_n = |\bv_0^\dagger\bG_n\bff_n|^2$ follows the exponential distribution with unit variance. Therefore, for $n\geq L$, the effective interference power from $T_n$ to $R_n$ is $I_n = r_n^{-\alpha}\rho_n$.

The effective power of the data link from $T_0$ to $R_0$ is given by $S = |\bv_0^\dagger\bH_0\bff_0|^2 = d^{-\alpha}|\bv_0^\dagger\bG_0\bff_0|^2$ with $S\geq \beta$ due to opportunistic transmission. Because the beamformers $\bv_0$ and $\bff_0$ are independent of $\bG_0$ as discussed in the preceding section, the random variable $W = |\bv_0^\dagger\bG_0\bff_0|^2$ has the exponential distribution with the following probability density function (PDF)
\begin{equation}\label{Eq:PDF:W}
f_W(w) = \exp(-w)/P_t, \quad w\geq \beta d^\alpha
\end{equation}
where  $P_t =\exp(-\beta d^\alpha)$.

\section{Outage Probability and Transmission Capacity}\label{Section:Outage:PerfectCSI}
Based on  auxiliary results in Section~\ref{Section:AuxResult}, bounds on  the SIR outage probability are obtained in Section~\ref{Section:Pout:Bounds:PerfCSI}. The scaling law of transmission capacity is derived for the regime of small outage probability in Section~\ref{Section:AsymTxCap:PerCSI}.

\subsection{Auxiliary Results}\label{Section:AuxResult}
To facilitate analysis, the interference nodes of $R_0$ after perfect interference cancelation, namely $\{T_n\mid n\geq L\}$, are separated into the strongest interferer $T_L$ and others $\{T_n\mid n\geq L+1\}$, referred to respectively as the \emph{primary} and the \emph{secondary} interferers. For convenience, denote the random interference power from $T_L$ as $G=I_L$. The  separation of interferers provides a useful result that conditioning on $G$, the secondary interferers $\{T_n\mid n\geq L+1\}$ form a Poisson point process as shown shortly. This  result is obtained by using the Marking Theorem \cite{Kingman93:PoissonProc}. To apply this theorem, a \emph{marked point process} is defined for the secondary interferers, where the mark of the node $T_n$ is the corresponding interference power $I_n$. Specifically, conditioning on the interference power of the primary interferer $G=g$, the desired marked point process is $\Pi(g) = \{(T_n, I_n\}\mid T_n\in \Phi/\{T_0\}, 0 \leq I_n < g\}$, where $\Phi$ is the transmitter process defined in Section~\ref{Section:NetworkModel}. Note that conditioning on $G=g$, the marks for different nodes are independent. Given this condition, the result in the following lemma directly follows from the Marking Theorem.
\begin{lemma}\label{Lem:MarkTheorem}\emph{The process $\Pi$ is a Poisson point process with the average number of nodes given by $\mu(g) = 2\pi\lambda\int\limits_0^\infty\int\limits_0^gr p(r,dI)dr$.
}
\end{lemma}
This result is useful for analyzing the aggregate interference from the secondary interferers to $R_0$. Conditioned on $G=g$, this interference is $I_{\Pi}(g) = \sum_{(T_n,I_n)\in\Pi^\star(g)}I_n$.

Next, the distribution of the interference power from both the primary ($G$) and the secondary interferers ($I_{\Pi}(g)$) are characterized in the following lemma.
\begin{lemma}\label{Lem:InterfPwr:l}\emph{The interference at $R_0$ has the following properties.
\begin{enumerate}
\item The PDF of the primary interference power $G$ is
\begin{equation}\label{Eq:PDF:G}
f_G(g) = \frac{\delta c_1^L \lambda^L g^{-\delta L -1}}{\Gamma(L)}\exp\left(-c_1\lambda g^{-\delta}\right)
\end{equation}
where $\delta = \frac{2}{\alpha}$ and $c_1 = \pi P_t\Gamma(\delta+1)$.
\item Conditioned on $G=g$, the mean and variance of the secondary interference power $I_{\Pi}(g)$ are
    \begin{eqnarray}
\E[I_\Pi(g)] &=&     \frac{2\pi\lambda\Gamma(\delta+1)}{\alpha - 2}g^{1-\delta}\\
\var(I_\Pi(g)) &=& \frac{\pi\Gamma(\delta+1)\lambda}{\alpha-1}g^{2-\delta}.
\end{eqnarray}
\end{enumerate}
}
\end{lemma}
\begin{proof}
See Appendix~\ref{App:InterfPwr:l}.
\end{proof}

\subsection{Bounds on  Outage Probability}\label{Section:Pout:Bounds:PerfCSI}
The analysis of the exact outage probability $\Pout(\lambda)  =  \E\left[\Pr(\SIR_0 \leq \theta)\right]$ is infeasible due to the difficulty in deriving the distribution function of the secondary interference power $I_\Pi(g)$. Therefore,
we resolve to obtaining bounds on $\Pout$ following the approach in \cite{WeberAndrews:TransCapAdHocNetwkDistSch:2006, WeberAndrews:TransCapWlssAdHocNetwkSIC:2005, WeberAndrews:TransCapWlssAdHocNetwkOutage:2005}. The outage probability can be written as
\begin{eqnarray}\label{Eq:Pout:Exact:a}
\!\!\!\Pout(\lambda)\!\!\!\!\!\! &=& \!\!\!\!\!\!\E\left[\Pr(I_{\Pi}(G) \geq W\theta^{-1}d^{-\alpha} - G \mid G, W) \mid WG^{-1}\!\!>  \right.\nn\\
&&\left.\theta d^\alpha\right]\Pr(WG^{-1}\geq \theta d^\alpha) + \Pr(WG^{-1}\leq \theta d^\alpha).
\end{eqnarray}
Thus,  a lower bound of $\Pout$  is given as
\begin{equation}
\Pout(\lambda) \geq \Pr(WG^{-1}\leq \theta d^\alpha).\label{Eq:PoutLB:PerfCSI}
\end{equation}
This lower bound considers only the primary interference, and hence is tight if the primary interferer $T_L$ is the dominant source of interference. Next, an upper bound of the outage probability can be derived by applying the following Chebyshev's inequality on \eqref{Eq:Pout:Exact:a}
\begin{equation}
\Pr(I_{\Pi}(g) \geq a ) \leq \min\left\{\frac{\var(I_\Pi(g))}{\left\{a -\E\left[I_\Pi(g)\right]\right\}^2}, 1\right\}\label{Eq:PoutUB:PerfCSI}.
\end{equation}
Based on \eqref{Eq:Pout:Exact:a}, \eqref{Eq:PoutLB:PerfCSI} and \eqref{Eq:PoutUB:PerfCSI}, bounds on  the outage probability are derived as shown in the following proposition.
\begin{proposition}\label{Prop:PoutBnds}\emph{For perfect CSI, the bounds on  the outage probability are given as follows.
\begin{enumerate}
\item The lower bound is
\begin{equation}\label{Eq:Pout:PerCSI:LB}
\Pout^L(\lambda) = \frac{1}{\Gamma(L)}\E\left[\gamma\left(L, c_2\lambda W^{-\delta}\right)\right]
\end{equation}
where  $c_2 = \pi\Gamma(1+\delta)\theta^\delta d^2$, and $W$ has the probability density function in \eqref{Eq:PDF:W}.
\item The upper bound is
\begin{equation}
\Pout^U(\lambda) = \Pout^L(\lambda) + \left[1 - \Pout^L(\lambda)\right]P_{\alpha}(\lambda) \label{Eq:Pout:PerCSI:UB}
\end{equation}
where
\begin{eqnarray}
P_{\alpha}(\lambda) \!\!\!\!\!\!&=& \!\!\!\!\!\E\left\{\left.\min\left[\frac{c_3\lambda G^{2-\delta}}{(d^{-\alpha}\theta^{-1}W-G-c_4 \lambda G^{2-\delta})^2}, 1\right] \right| \right.\nn\\
&&\left.\frac{W}{G}> d^\alpha\theta\right\}\label{Eq:Prob:Comp}
\end{eqnarray}
and $c_3 = \frac{2\pi\Gamma(\delta+1)}{\alpha-2}$ and $c_4 = \frac{\pi\Gamma(1+\delta)}{\alpha-1}$.
\end{enumerate}
}
\end{proposition}
The proof is straightforward and is hence omitted. The difficulty in deriving a closed-form expression for $\Pout$ is mainly lies in that the distribution function of $I_\Pi(G)$, called a \emph{shot noise process}, is unknown \cite{Lowen:PowerLawShotNoise:1990, WeberAndrews:TransCapAdHocNetwkDistSch:2006}.

\subsection{Asymptotic Transmission Capacity}\label{Section:AsymTxCap:PerCSI}
In this section, the scaling law for transmission capacity is derived for small target outage probability ($\epsilon\rightarrow 0$). This scaling law also accurately characterizes transmission capacity in the non-asymptotic outage regime (up to 0.1) as shown by simulations in Section~\ref{Section:Simualtion}.

Small target outage probability results in a network of sparse transmitting nodes (i.e. $\lambda\rightarrow 0$). For such a sparse network, the useful relationship between the outage probability and node density is derived and shown in the following lemma.
\begin{lemma}\label{Lem:AsymPout:PerfCSI}\emph{
For $\lambda\rightarrow 0$, the outage probability scales with $\lambda$ as follows.
\begin{enumerate}
\item For $L\leq \alpha$,
\begin{equation}
\kappa_1 \leq \lim_{\lambda\rightarrow 0 }\frac{\Pout(\lambda)}{\lambda^L} \leq  \kappa_1(1 + \kappa_2)
\end{equation}
where $\kappa_1 = \frac{\Gamma(1-\delta L, \beta d^\alpha)[\pi\Gamma(\delta+1)\theta^\delta d^2]^L}{P_t\Gamma(L+1)}$ and $\kappa_2 = \frac{2^{\delta L }}{L}\left(\frac{\alpha}{\alpha-2}-2^{-\delta}\right)$.
\item For $L > \alpha$,
\begin{equation}
\kappa_1 \leq \lim_{\lambda\rightarrow 0 }\frac{\Pout(\lambda)}{\lambda^L}, \quad \lim_{\lambda\rightarrow 0 }\frac{\Pout(\lambda)}{\lambda^{\alpha}} \leq  \kappa_3
\end{equation}
where $\kappa_3 = \frac{8(c_1d^{2}\theta^\delta)^\alpha\Gamma(-1, \beta d^\alpha)\Gamma(L-\alpha+1)}{(\alpha-2)P_t\Gamma(L)}$.
\end{enumerate}
}
\end{lemma}
\begin{proof}
See Appendix~\ref{App:AsymPout:PerfCSI}.
\end{proof}

Using Lemma~\ref{Lem:AsymPout:PerfCSI} and the definition of transmission capacity  in \eqref{Eq:TxCap}, the main result of this section is obtained and summarized in the following theorem.
\begin{theorem}\label{Theo:TxCap}\emph{
For small target outage probability $\epsilon \rightarrow 0$, the transmission capacity scales as
\begin{enumerate}
\item For $L\leq \alpha$,
\begin{equation}
\lim_{\epsilon\rightarrow 0}\frac{C(\epsilon)}{[\kappa_1(1+\kappa_2)]^{-\frac{1}{L}}\epsilon^{\frac{1}{L}}} \geq 1, \quad \lim_{\epsilon\rightarrow 0}\frac{C(\epsilon)}{\kappa_1^{-\frac{1}{L}}\epsilon^{\frac{1}{L}}} \leq 1
\end{equation}
where $\kappa_1$ and $\kappa_2$ are specified in Lemma~\ref{Lem:AsymPout:PerfCSI}.
\item For $L > \alpha$,
\begin{equation}
\lim_{\epsilon\rightarrow 0}\frac{C(\epsilon)}{\kappa_3^{-\frac{1}{\alpha}}\epsilon^{\frac{1}{\alpha}}} \geq 1, \quad \lim_{\epsilon\rightarrow 0}\frac{C(\epsilon)}{\kappa_1^{-\frac{1}{L}}\epsilon^{\frac{1}{L}}} \leq 1
\end{equation}
where $\kappa_3$ is given in Lemma~\ref{Lem:AsymPout:PerfCSI}.
\end{enumerate}
}
\end{theorem}
The above theorem shows that as the target outage probability decreases, transmission capacity grows following the power law  $a\epsilon^t$ where $a$ and $t$ are constants. For $L > \alpha$, only bounds of the exponent $t$ are known. The derivation of the exact exponent may require analyzing the distribution function of the secondary interference power (cf. Section~\ref{Section:EffModel:PerfCSI}) rather than using Chebyshev's inequality  in \eqref{Eq:PoutUB:PerfCSI}. Unfortunately, no closed-form expression for this distribution function is known for the present case \cite{Lowen:PowerLawShotNoise:1990}.

For $L \leq \alpha$, the exponent of the transmission capacity power law $a\epsilon^t$ is shown in Theorem~\ref{Theo:TxCap} to be $t = 1/L$, and $\alpha$ is bounded as $[\kappa_1(1+\kappa_2)]^{-\frac{1}{L}} \leq \alpha \leq \kappa_1^{-\frac{1}{L}}$. This power law indicates that the size of antenna array $L$ determines the sensitivity of transmission capacity to the change on the outage constraint. To facilitate our discussion, rewrite the scaling law in Theorem~\ref{Theo:TxCap} as $C(\epsilon) \cong \alpha \epsilon^{\frac{1}{L}}$ where ``$\cong$" represents asymptotic equivalence for $\epsilon \rightarrow 0$. Moreover, consider two sets of values $(C_1,\epsilon_1)$ and $(C_2, \epsilon_2)$, and define the logarithmic ratios $\Delta C = \log\frac{C_1}{C_2}$ and $\Delta \epsilon = \log\frac{\epsilon_1}{\epsilon_2}$. Using this notation, the above scaling law can be simplified as
\begin{equation}\label{Eq:TxCap:Sensitive}
\frac{\Delta C}{\Delta \epsilon} \cong \frac{1}{L}.
\end{equation}
The above quantity $\frac{\Delta C}{\Delta \epsilon}$ represents the sensitivity of transmission capacity towards the change of the outage constraint. Its value decreases inversely with the size of antenna array. Specifically, computed using \eqref{Eq:TxCap:Sensitive}, a hundred-time decrease on $\epsilon$ reduces network transmission capacity by $\{10, 3.2, 1.8\}$ times for $L=\{2, 4, 8\}$, respectively.

\section{Simulation and Discussion}\label{Section:Simualtion}
The results in this section are obtained by simulating a MANET following the procedure in \cite{WeberKam:CompComplexMANETs:2006}.

\begin{figure}
\centering
\vspace{-10pt}\includegraphics[width=9cm]{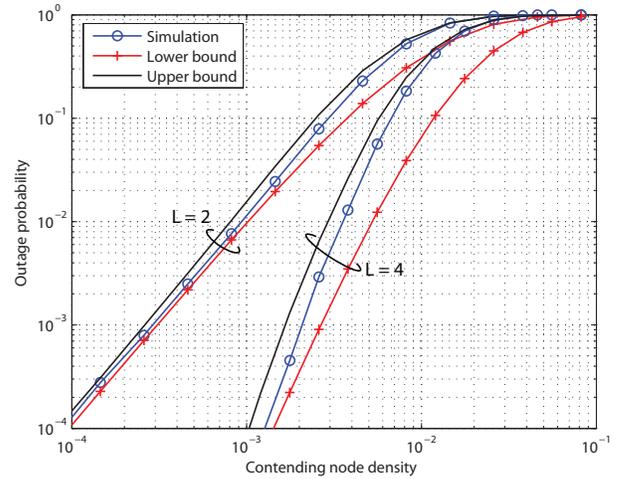}\\
\vspace{-10pt}  \caption{Outage probability for different  transmitting node densities and perfect CSI.  The size of the antenna array is $L=\{2, 4\}$. }\label{Fig:Pout:PerfCSI}
\end{figure}

For perfect CSI, the bounds on  outage probability from Proposition~\ref{Prop:PoutBnds} and simulated values are compared in Fig.~\ref{Fig:Pout:PerfCSI}. The path loss exponent is $\alpha = 4$ and the number of antenna per node is $L=\{2,4\}$. Two  observations can be made from Fig.~\ref{Fig:Pout:PerfCSI}. First, the bounds for $L=2$ are tighter than those for $L=4$. Second, the bounds and  the simulated values of the outage probability converge as the transmitting node density $\lambda$ decreases. These two observations can be explained by the dominance of the primary interference over the secondary one  as $L$ increases or $\lambda$ decreases, where the secondary interference causes the looseness of the bounds on outage probability.

\begin{figure}
\centering
\vspace{-10pt}\includegraphics[width=9cm]{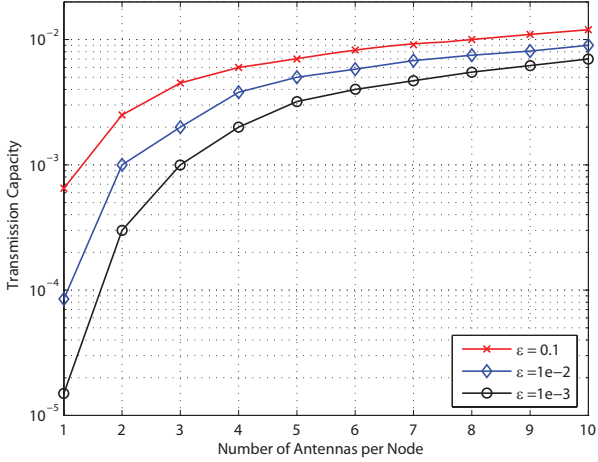}\\
\vspace{-10pt}  \caption{Transmission capacity by simulation for different  node densities and perfect CSI. The  size of the antenna array is $L=4$ and the outage constraint is $\epsilon = \{10^{-1},10^{-2}, 10^{-3}\}$. }\label{Fig:TxCap:Eps}
\end{figure}

In Fig.~\ref{Fig:TxCap:Eps}, the transmission capacity is plotted for an increasing number of antennas per node assuming perfect CSI. Furthermore, different outage constraints, namely $\epsilon = \{10^{-1}, 10^{-2}, 10^{-3}\}$, are considered. From Fig.~\ref{Fig:TxCap:Eps}, the following observations are made. First, the use of multiple antennas for interference cancelation leads to the increase in transmission capacity by an order of magnitude or more with respect to the case of single-antenna per node. This capacity gain is especially large for a small number of antennas and small target outage probability. For example, for $\epsilon = 10^{-1}$, the use of three antennas per node provides transmission capacity seven times of that for the single-antenna case. The capacity gain by using additional antennas diminishes rapidly as the number of antennas per node increases. Second, the outage constraint affects transmission capacity significantly for a small number of antennas per node. Nevertheless, transmission capacity becomes insensitive to the change on the outage constraint as the number of antennas increases.

\begin{figure}
\centering
\vspace{-10pt}\includegraphics[width=9cm]{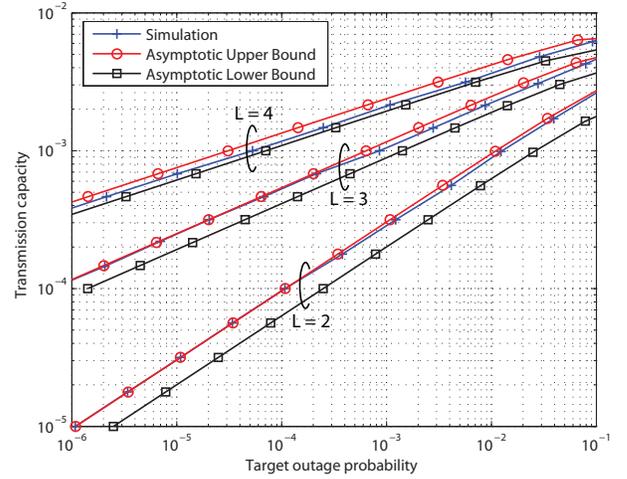}\\
\vspace{-10pt}  \caption{Comparison between asymptotic bounds on transmission capacity and the exact values obtained by simulation perfect CSI and the size of antenna array $L=\{2, 3, 4\}$}\label{Fig:TxCapScale}
\end{figure}

In Fig.~\ref{Fig:TxCapScale}, asymptotic bounds on transmission capacity in Theorem~\ref{Theo:TxCap} are compared with the exact values obtained by simulation for perfect CSI and the range of target outage probability  $\epsilon \in [10^{-5}, 0.1]$. As observed from  Fig.~\ref{Fig:TxCapScale}, the asymptotic upper bound on transmission capacity is very tight for $L=\{2, 3\}$ even in the non-asymptotic range e.g. $\epsilon \in [0.01, 0.1]$. The tightness of this bound is due to the dominance of primary interference for interference cancelation with small sizes of antenna array. For $L=4$, both the asymptotic lower and upper bounds are tight. The tightness of asymptotic bounds implies the slopes of the transmission capacity vs. target outage probability curves are approximately  equal to $\frac{1}{L}$.

\appendix

\subsection{Proof of Lemma~\ref{Lem:InterfPwr:l}}\label{App:InterfPwr:l}
Define a marked Poisson point process as $\mathcal{M}(g) = \{(T_n, I_n\}\mid T_n\in \Phi/\{T_0\}, I_n \geq g\}$. Given that  $I_n = r_n^{-\alpha}\rho_n$, the node density of $\mathcal{M}(g)$ follows from the Marking Theorem
\begin{eqnarray}
\mu(\mathcal{M}(g)) &=& \lambda\int_0^\infty\int_0^{(t/g)^{1/\alpha}} 2\pi r f_\rho(t)drdt\nn\\
&=& \pi \Gamma(\delta+1)\lambda g^{-\delta}, \quad \delta := \frac{2}{\alpha}\nn.
\end{eqnarray}
The cumulative density function (CDF) of $G$ is the probability that the number of nodes in the subset $\mathcal{M}(g)$ is no more than $L-1$. Thus,
$
\Pr(G\leq g) = \sum_{k=0}^{L-1}\frac{\left(c_1 \lambda g^{-\delta}\right)^k}{k!}\exp\left(-c_1\lambda g^{-\delta}\right).
$
The probability density function of $G$ is obtained by differentiating the above CDF.

Using Campbell's theorem\cite{Kingman93:PoissonProc}, the expressions for $\var(I_\Pi(g))$ and $\E[I_\Pi(g)]$ are obtained as
\begin{eqnarray}
\E[I_\Pi\mid g] &=& 2\pi\lambda\int_0^\infty \int_{\left(\frac{\rho}{g}\right)^{1/\alpha}}^\infty r^{1-\alpha}\rho e^{-\rho} dr d\rho\nn\\
&=& \frac{2\pi\lambda\Gamma(\delta+1)}{\alpha - 2}g^{1-\delta}.
\end{eqnarray}
\begin{eqnarray}
\var(I_\Pi\mid g) &=& 2\pi\lambda\int_0^\infty\int_{\left(\frac{\rho}{g}\right)^{1/\alpha}}^\infty  r\left(r^{-\alpha}\rho\right)^2 e^{-\rho} d\rho dr\nn\\
&=& \frac{\pi\Gamma(\delta+1)\lambda}{\alpha-1}g^{2-\delta}.
\end{eqnarray}

\subsection{Proof of Lemma~\ref{Lem:AsymPout:PerfCSI}}\label{App:AsymPout:PerfCSI}
{\bf Lower Bound:}
By expanding \eqref{Eq:Pout:PerCSI:LB} around $\lambda = 0$ based on Taylor's series, it can be shown that
\begin{equation}\label{Eq:AsymPout:App2}
\Pout^L(\lambda) = \kappa \lambda^L + O(\lambda^{L+1})
\end{equation}
where $\kappa$ is given in Lemma~\ref{Lem:AsymPout:PerfCSI}.

{\bf Upper Bound:}
The asymptotic expansion of the upper bound in \eqref{Eq:Pout:PerCSI:UB} is obtained as follows. To simplify notation, define $B = d^{-\alpha}\theta^{-1}W$. The term in \eqref{Eq:Prob:Comp} can be written as $P_{\alpha}(\lambda) = \E[\Lambda(B)]$ where the function $\Lambda(b)$ is defined as
\begin{equation}
\Lambda(b) = \E\left\{\left.\Omega \right| G < b\right\}\Pr(G < b).\nn
\end{equation}
where $\Omega = \min\left[\frac{c_3\lambda G^{2-\delta}}{(b-G-c_4 \lambda G^{2-\delta})^2}, 1\right]$.
It can be expanded as
\begin{eqnarray}
\!\!\!\Lambda(b) \!\!\!&=&\!\!\!\! \E\left\{\left. \Omega\right| G \leq \frac{b}{2}\right\}\Pr\left(G \leq \frac{b}{2} \right)+\E\left\{\left.\Omega\right| \frac{b}{2} < G < b\right\}\times \nn\\
&&\Pr\left(\frac{b}{2} < G < b\right)\nn\\
%===========================
&\leq& \!\!\!\! \E\left\{\left. \Omega\right| G \leq \frac{b}{2}\right\}\Pr\left(G \leq \frac{b}{2}\right)+ \Pr\left(\frac{b}{2} < G < b\right)\nn
\end{eqnarray}
\begin{eqnarray}
%==========================
&\leq& \!\!\! 4c_3\lambda b^{-2}\left[1+O(\lambda)\right]\underbrace{\E\left[G^{2-\delta}\mid G \leq \frac{b}{2} \right]\Pr\left(G \leq \frac{b}{2} \right)}_{\Lambda_1(b)} + \nn\\
&&\underbrace{\Pr\left(\frac{b}{2} < G < b\right)}_{\Lambda_2(b)}, \quad \lambda \rightarrow 0.\label{Eq:App:a}
\end{eqnarray}
Next, by using the probability density function in \eqref{Eq:PDF:G},
\begin{equation}
\Lambda_1(b) = \frac{(c_1\lambda)^{\alpha-1}}{\Gamma(L)}\int_{c_1\lambda(b/2)^{-\delta}}^\infty g^{L-\alpha}\exp(-g)d g. \label{Eq:App:b}
\end{equation}
Note that for $\lambda \rightarrow 0$, the integral in \eqref{Eq:App:b} is bounded for $L > \alpha$ and unbounded for $L \leq \alpha$. By separating these two cases,
\begin{equation}
\left\{\begin{aligned}
\Lambda_1(b)&= \frac{(c_1\lambda)^{\alpha-1}}{\Gamma(L)}\Gamma\left(L-\alpha+1, c_1\lambda\left(b/2\right)^{-\delta}\right), &&L >  \alpha \\
\Lambda_1(b)&\leq \frac{c_1^{L-1}\lambda^{L-1}(b/2)^{-\delta(L-\alpha)}}{\Gamma(L)}[1+O(\lambda)],&&L \leq \alpha.
\end{aligned}\right.
\end{equation}
Again, by using \eqref{Eq:PDF:G},
\begin{eqnarray}
\Lambda_2(b) &=& \frac{1}{\Gamma(L)}\int^{c_1\lambda(b/2)^{-\delta}}_{c_1\lambda b^{-\delta}} g^{L-1}\exp(-g)d g\nn\\
&\leq& \frac{[c_1\lambda(b/2)^{-\delta}]^{(L-1)}}{\Gamma(L)}\left[e^{-c_1\lambda b^{-\delta}}-e^{-c_1\lambda(b/2)^{-\delta}}\right]\nn\\
&\leq& \frac{2^{\delta L}(1-2^{-\delta})c_1^{L}\lambda^{L}b^{-\delta L}}{\Gamma(L)} + O(\lambda^{L+1}). \label{Eq:App:c}
\end{eqnarray}
By combining \eqref{Eq:App:c}, \eqref{Eq:App:b} and \eqref{Eq:App:a},
\begin{equation}
\E[\Lambda(B)]\leq \left\{\begin{aligned}
&\frac{4\E[B^{-2}]c_3 c_1^{\alpha-1}\Gamma\left(L-\alpha+1, c_1\lambda\left(b/2\right)^{-\delta}\right) }{\Gamma(L)}\lambda^\alpha + \\
&O(\lambda^{\alpha+1}),\quad L > \alpha\\
&\frac{\left(\frac{\alpha}{\alpha-2}-2^{-\delta}\right)2^{\delta L}c_1^L\E[B^{-\delta L}]}{\Gamma(L)}\lambda^L+O(\lambda^{L+1})\\
& L \leq \alpha
\end{aligned}\right.\label{Eq:App:d}
\end{equation}
where by using \eqref{Eq:PDF:W}
\begin{eqnarray}
\E[B^{-\delta L}] &=& \frac{d^{2L}\theta^{\delta L}}{P_t}\Gamma(1-\delta L, \beta d^\alpha)\nn\\
\E[B^{-2}] &=& \frac{d^{2\alpha}\theta^2\Gamma(-1, \beta d^\alpha)}{P_t}.
\end{eqnarray}
The desired upper bound is obtained by combining \eqref{Eq:Pout:PerCSI:UB}, \eqref{Eq:AsymPout:App2} and \eqref{Eq:App:d}.

\section*{Acknowledgment}
K. Huang is the recipient of the University Continuing Fellowship from The University of Texas at Austin. This work is funded by the DARPA IT-MANET program under the grant W911NF-07-1-0028.

\bibliographystyle{ieeetr}

\end{document}